\documentclass[]{llncs}
\usepackage[T1]{fontenc}
\usepackage{graphicx}
\usepackage{amsthm}
\theoremstyle{definition}

\usepackage{amsmath}
\usepackage{setspace}
\usepackage{caption} 
\usepackage{csquotes}
\usepackage{booktabs}
\usepackage{tabularx}
\usepackage{listings}
\usepackage{comment}
\usepackage{enumerate}
\usepackage{float}
\usepackage{xcolor}
\usepackage{nameref}
\usepackage{hyperref}
\usepackage{multirow}
\usepackage{makecell}
\usepackage{diagbox}
\usepackage{algorithm2e}
\RestyleAlgo{ruled}
\usepackage{wrapfig}
\usepackage{array}
\newcolumntype{P}[1]{>{\centering\arraybackslash}p{#1}}

\usepackage{todonotes}

\begin{document}

\title{Privacy and Confidentiality Requirements Engineering for Process Data}

%
\author{Fabian Haertel\inst{1}, Juergen Mangler\inst{1}, Nataliia Klievtsova\inst{1}, Celine Mader\inst{2}, Eugen Rigger\inst{2}, Stefanie Rinderle-Ma\inst{1}}


\institute{Technical University of Munich, Garching, Germany\\ TUM School of Computation, Information, and Technology \\ 
\email{\{firstname.lastname\}@tum.de}\\~\\
\and Zumtobel Lighting GmbH, Dornbirn, Austria,\\ \email{\{firstname.lastname\}@zumtobelgroup.com}}
%
\maketitle              
\begin{abstract}
The application and development of process mining techniques face significant challenges due to the lack of publicly available real-life event logs. One reason for companies to abstain from sharing their data are privacy and confidentiality concerns. Privacy concerns refer to personal data as specified in the GDPR and have been addressed in existing work by providing privacy-preserving techniques for event logs. However, the concept of confidentiality in event logs not pertaining to individuals remains unclear, although they might contain a multitude of sensitive business data. This work addresses confidentiality of process data based on the privacy and confidentiality engineering method (PCRE). PCRE interactively explores privacy and confidentiality requirements regarding process data with different stakeholders and defines privacy-preserving actions to address possible concerns. We co-construct and evaluate PCRE based on structured interviews with process analysts in two manufacturing companies. PCRE is generic, hence applicable in different application domains. The goal is to systematically scrutinize process data and balance the trade-off between privacy and utility loss. 

\keywords{Process mining, Privacy and Confidentiality Requirements, Process data}
\end{abstract}

\section{Introduction}
\label{sec:intro}

Gartner estimates that the process
mining market will grow to $2.3$ billion by 2025\footnote{\url{https://www.gartner.com/en/documents/4007520}, last access: 2025-01-25}. Uber, for example, achieved $20$M in efficiency gains by optimizing their processes \cite{Rowlson2020}. However, there is a significant need to address privacy and confidentiality in process mining as organizations and companies often hesitate to share information due to competitive reasons \cite{DBLP:conf/bpm/AalstAM11}. This refers to sharing event log data for analysis purposes as well as sharing event logs in inter-organizational settings \cite{DBLP:conf/caise/ElkoumyFDLPW20}. The latter hinders building process collaborations and networks. 

Privacy refers to rights of individuals regarding the collection, usage, and disclosure of their personal data as specified in regulations such as the GDPR\footnote{\url{https://gdpr-info.eu/}, last access: 2025-01-25}. Confidentiality \textsl{``is concerned with data access, whereas privacy is focused on individuals and their
rights. When the data are personal data, the confidentiality challenges coincide with the privacy
challenges''.} \cite{DBLP:journals/tmis/ElkoumyFSKMVRW22}. 
Process data sharing under privacy and confidentiality concerns, necessitates the development of privacy and confidentiality preserving techniques. This is also underpinned by the research line of responsible process mining committing to the FACT criteria of Fairness, Accuracy, Confidentiality, Transparency \cite{DBLP:books/sp/22/Mannhardt22}. 

Privacy in process mining has been investigated by various approaches in the last years, providing privacy models and privacy-preserving techniques for event logs, e.g., \cite{DBLP:journals/dke/FahrenkrogPetersenAW23,DBLP:journals/is/ElkoumyPD23}, also for privacy-sensitive domains such as healthcare \cite{DBLP:conf/bpm/PikaWBHAR19}. Here, the GDPR gives clear guidance on protection of
personal information of individuals. However, on top of individual data, process event logs might contain confidential data \cite{DBLP:journals/istr/BakhtinaMS23,DBLP:journals/sttt/DumasGJLMPPPTTY22}. The severity of disclosing confidential information varies between organizations and industries.  Revealing a detailed production
processes and quality control steps in the manufacturing sector could expose business secrets or
operational weaknesses. In the financial industry, providing information about the process steps
for fraud detection algorithms or loan approval processes could provide valuable information to
competitors. Healthcare analysts point out possible privacy and confidentiality concerns related to time stamps in event logs. Until now, no generally applicable approach exists to explore an organization’s privacy
and confidentiality requirements in a systematic and interactive way and, at the same time, to balance usefulness of the process analysis with privacy and confidentiality requirements.

Hence, in this work, we propose the privacy and confidentiality requirements engineering method (PCRE). The goal is to scrutinize process data regarding privacy and confidentiality concerns in a systematic way with different stakeholders and define possible mitigation actions such as data anonymization. At this, the goal is to balance privacy and utility loss \cite{DBLP:journals/tmis/ElkoumyFSKMVRW22}. PCRE is co-constructed and evaluated based on structured interviews with process analysts in two manufacturing companies and additionally evaluated on an ideation process. It consists of seven phases with interview templates, checklists, and guidelines. The output is a consolidated executive summary containing the privacy and confidentiality assessment of all process elements and data, together with mitigation actions. Doing so, PCRE aims at supporting organizations and companies to enable the publication of process event logs with confidentiality assurance. This paper is based on the first author’s Masters thesis \cite{Haertel24} and structured as follows: PCRE is presented in Sect. \ref{sec:pcre} and evaluated in Sect. \ref{sec:eval}. Section \ref{sec:relwor} discusses related work. Section \ref{sec:conclusion} concludes the paper.

\section{Privacy and Confidentiality Engineering Method}
\label{sec:pcre}

The privacy and confidentiality engineering method (PCRE) aims at systematically collecting the privacy and confidentiality requirements that are relevant for a given process and its stakeholders. Moreover, PCRE aims at determining how the derived  requirements can be met by applying privacy-preserving techniques. This results in the following objectives for PCRE harvested based on experience from collaborations with one hospital and one manufacturing company:

\begin{itemize}
    \item \textsl{Objective 1:} Define process data usage objectives, e.g., company-internal or public analysis of the process data and scope of contained data.
\item	\textsl{Objective 2:} Outline the benefits of publishing process data, i.e., what does the company gain in exchange? ($\mapsto$ balance between utility of analysis results and privacy and confidentiality requirements)
\item	\textsl{Objective 3:} Scrutinize process elements systematically for necessary data anonymization tasks in order to preserve privacy and confidentiality.
\item	\textsl{Objective 4:} Align the data usage objectives with the necessary data anonymization steps in order to balance data utility and privacy loss.
\end{itemize}

\begin{figure}[htb!]
    \centering
    \includegraphics[width=\textwidth]{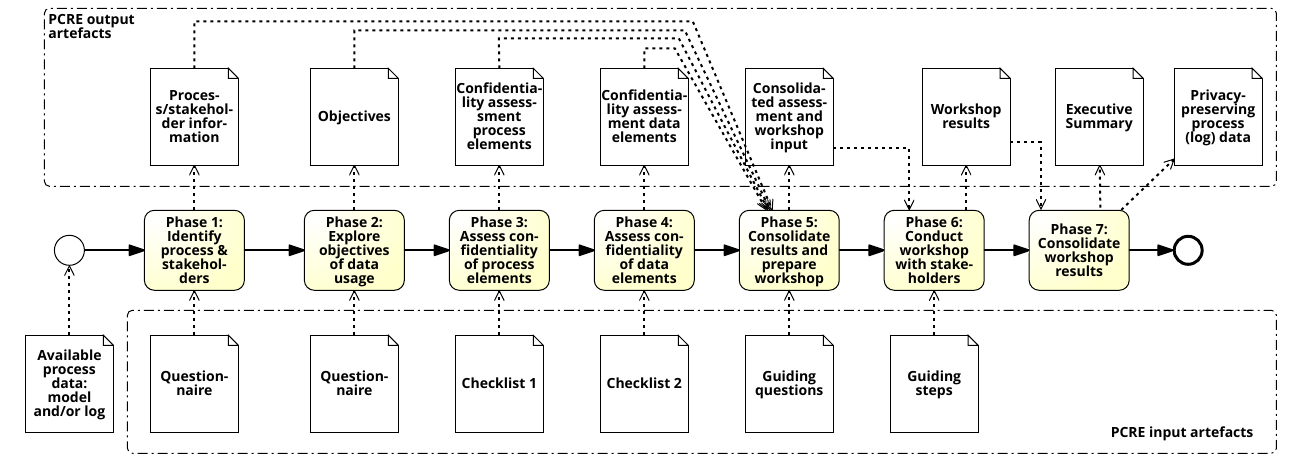}
     \caption{PCRE Method (modeled as BPMN process using SAP Signavio)}
    \label{fig:pcre_method}
\end{figure}

PCRE takes as \textbf{input} available process data (cf. Fig. \ref{fig:pcre_method}), i.e., process models or event logs. An important success factor to assess privacy and confidentiality is the availability of information about process data. If, for example, an event log contains control flow related information, PCRE can assess tasks, their order, and their timestamps. In the real-world manufacturing case considered in this paper, the PCRE assessment is based on a process model that is implemented and executed through the process engine CPEE\footnote{\url{cpee.org}}. As the model is executable, it has fully specified data flow and endpoints for service invocation. Moreover, the logging mechanism of the CPEE stores all execution information in the event logs, including data element values and endpoint information. Hence, PCRE assessment of the process model corresponds to privacy and confidentiality assessment of the generated event logs. PCRE can also be applied to event logs without an underlying process model. This might especially useful for object-centric event logs\footnote{\url{https://ocel-standard.org/}} as they are built around data. One benefit of assessing process models is that they typically provide a full overview on process behavior whereas logs reflect the behavior that has been observed, but might miss unseen behavior caused by, e.g., infrequent traces \cite{chen2023predicting}. For a more detailed discussion on process behavior reflected by models, logs, and systems, we refer to \cite{Aalst_Relating_2018}. PCRE can be applied to newly designed or re-designed process models during design time. It can also be applied to event logs ex post or during runtime. If helpful for stakeholders, a process model might be additionally discovered from the logs. Again this might be especially useful for object-centric logs as they might represent more than one process model.  

The output of PCRE comprises an executive summary of privacy and confidentiality requirements and the process event log that has been prepared for sharing with privacy-preserving techniques determined during PCRE assessment .

The PCRE aims to meet \textsl{Objective 1} to \textsl{Objective 4} by systematically scrutinizing process data with different stakeholders along well-defined phases and input artefacts such as interview guidelines. It consists of seven phases as shown in Fig. \ref{fig:pcre_method}, including the PCRE artefacts and the involved stakeholders. PCRE was initially designed by the authors based on experience and then refined in a first interview round with process analysts from the manufacturing domain. One design goal of PCRE is to follow a bottom-up approach, i.e., a more detailed confidentiality assessment of process structure and process data is conducted with process and business analysts in the beginning and then abstracted for discussion with other stakeholders, including the legal compliance team.

\noindent\textbf{PCRE Phase 1: Identify process \& stakeholders (Stakeholders: Process Analyst, Management)} Phase 1 aims to collect general knowledge for shaping and conducting the subsequent phases. Phase 1 involves process analysts and optionally the management for additional insights. Table \ref{tab:phase1} provides the template for the written questionnaire as artefact. 
Question 1.1 determines which process information is available, i.e., process event logs with or without a process model. Note that a process model can be discovered from the event logs. This model reflects the behavior stored in the logs. If an additional process model is available, this model can provide additional information when compared to the logs, including additional behavior or additional information on process data. The latter is the case, if the process model is not implemented and executed, but modeled by, e.g., domain experts.
Question 1.1 also asks which process mining task such as conformance checking or predictions is envisioned and possible and which data preparation steps are required. 
Different process mining tasks might bear different confidentiality requirements. For conformance checking, the detection of alignments might reveal sensitive business secrets, e.g., if a certain process task takes always longer than specified in the model or if a certain data value such as temperature exceeds a certain threshold several times. Data preparation steps can also be source for confidentiality requirements and possible action points for privacy-preserving techniques. If, for example, log information from different data sources is merged when generating process logs, business-sensitive data might be revealed. Questions 1.2 and 1.3 provide insights into data analysis practices and usage of the data ($\mapsto$ \textsl{Objective 1}). 
which departments are using the data. Question 1.4 identifies a list of stakeholders relevant to the analysis of privacy
and confidentiality requirements within the organization and a collection of essential information to
guide the subsequent phases. 

\begin{table}[htb!]
\begin{scriptsize}
 \centering\begin{tabular}{p{0.5cm}p{11cm}}
\textbf{ID} & \textbf{Question}   \\\hline
1.1 & Available process information: i) process model and process event logs, ii) process event logs  $\mapsto$ process mining task, data preparation steps  \\
1.2 & Process data usage: is the process data currently used for data analysis? If so, to what extent? \\
1.3 & Process data usage: Which departments are currently working with the process data, and who uses the
analysis results?  \\
1.4 & Which stakeholders are relevant for the subsequent phases?  \\\hline
\end{tabular}\caption{Template for (written) questionnaire in Phase 1}\label{tab:phase1}
\end{scriptsize}
\end{table}
\vspace{-0.5cm}

\noindent\textbf{PCRE Phase 2: Explore objectives of data usage (Stakeholders: Process Analyst, Management, Business Analyst, Additional Stakeholders, e.g., finance department, human resources, and operations)} Phase 2 is conducted through  interviews to increase the depth of interaction, influencing the richness of insights. Table \ref{tab:phase2} summarizes six guiding questions for assessing the usage of process data ($\mapsto$ \textsl{Objective 1}) and the actual/possible objectives of the usage. Questions 2.1--2.6 can be applied based on a process model or an event log depending on the available information (signified by an `x' in Tab. \ref{tab:phase2}).

\begin{table}[htb!]
\begin{scriptsize}
\centering\begin{tabular}{p{0.5cm}p{9.3cm}P{1cm}P{1cm}}
\textbf{ID} & \textbf{Question} & \textbf{Model} & \textbf{Log}\\\hline
2.1 & Is process data currently utilized for analysis? & x & x\\
2.2 & Which aspects of the process analysis are of interest?  & x & x\\
2.3 & Which analysis results are considered useful?  & x & x\\
2.4 & Which metrics/KPIs, e.g., duration or frequency of process tasks, are currently calculated for process analysis? How are they calculated?  & x & x\\
2.5 & Which metrics/KPIs are currently not calculated for process analysis? Which of them would be of interest? Why are they (currently) not calculated?  & x & x\\
2.6 & Which insights from process analysis process would help to manage the process more effectively for each department?  & x & x\\\hline
\end{tabular}\caption{Template for interview-based questionnaire in Phase 2}\label{tab:phase2}
\end{scriptsize}
\end{table}
\vspace{-0.5cm}
Questions 2.1 -- 2.6 aim to consolidate the process analysis requirements across different departments, also by including additional stakeholders. The collected and consolidated process analysis requirements can then be used for evaluating the utility loss by data anonymization. Furthermore, the inclusion and ``sharing'' of insights from the responses might increase the motivation of stakeholders to participate in PCRE. Questions 2.1 to 2.6 explore different facets of process data usage, including its current application for analysis (2.1), the specific interests and aspects of the process analysis that stakeholders find compelling (2.2), and the types of results considered useful from such analyses (2.3). They explore the metrics and KPIs currently used for process analysis, their calculation methods (2.4), and any potentially valuable metrics/KPIs not yet calculated, including reasons for their omission (2.5). Additionally, this question aim to identify areas for improvement. Finally, stakeholders are asked about the process insights that could enhance department-specific management precision (2.6) ($\mapsto$ \textsl{Objective 2}).

\noindent\textbf{PCRE Phase 3: Assess confidentiality of process elements (Stakeholders: Process Analyst, Management, Additional Stakeholders)} process analysts and management participate in a collaborative review session to thoroughly analyse the confidentiality of the process metadata ($\mapsto$ \textsl{Objective 3}). 
\begin{table}[htb!]
\begin{scriptsize}
\centering\begin{tabular}{p{0.5cm}p{9.8cm}P{0.8cm}P{0.7cm}}
\textbf{ID} & \textbf{Question} & \textbf{Model} & \textbf{Log}\\\hline
3.1 & For all existing subprocesses, are there any representing, unique or innovative business procedures, i.e., their disclosure could harm the organization? & x & \\
3.2 & For all considered processes and subprocesses, do their names directly or indirectly indicate confidential business areas, technologies, or methods that should not be public? & x & $\sim$\\
3.3 & For all tasks within the (sub)process under consideration, are there tasks that represent unique or innovative business procedures, i.e., their disclosure could harm the organization?\\
3.4 &For all tasks within the (sub)process under consideration, could the task label reveal specific business activities or proprietary services that should be protected for competitive reasons? & x & x \\
3.5 &For all process parameters, including timestamps, within the (sub)process under consideration, do they contain sensitive data such as customer-specific information or operational settings that should not be publicly accessible? & x  & x \\
3.6 &For all endpoints within the (sub)process under consideration, could the disclosure of these endpoints provide details about the organization’s network infrastructure, security mechanisms, or external service providers? &  $\sim$ & $\sim$\\
3.7 &Does the documentation of the process contain specific information about the reasons for changes that could indicate business strategies or product developments? & \multicolumn{2}{c}{$^*$} \\\hline
\multicolumn{4}{l}{$\sim$: if information is available, i.e., for executable models, for logs containing data values;}\\
 \multicolumn{4}{l}{$^*$: if documentation, e.g., a change log, is available}
\end{tabular}\caption{Checklist template for Phase 3}\label{tab:phase3}
\end{scriptsize}
\end{table}
\vspace{-0.5cm}
Questions 3.1 to 3.7 assess privacy and confidentiality concerns, cf. Tab \ref{tab:phase3}. For Question 3.1, a process model containing subprocesses is required. In this case, models discovered from event logs are not suitable as typically they do not contain subprocesses. If no subprocess exists, Question 3.1 is skipped. If subprocesses is exist, they are examined to identify confidential subprocedures that could potentially reveal specific business functions or strategies that may interest competitors or external parties. Subprocesses can also provide insights into proprietary methods or technologies that the company considers a competitive advantage. Question 3.2 considers the names of processes and subprocesses. Here, the focus is on determining whether these labels expose sensitive business strategies or technologies. Question 3.3 examines the existence of tasks within the process that could reveal confidential operational functions or innovative strategies. Accordingly, Question 3.4 considers how task labels  might reflect business logic or unique services. Question 3.5 scrutinizes process parameters to ascertain whether they hold sensitive data. This also includes timestamps of tasks in logs as they could reveal sensitive information, e.g., about treatments of patients. Question 3.6 aims at endpoints as their disclosure might provide hints to the IT infrastructure, potentially revealing vulnerabilities that could be exploited by adversaries. Finally, Question 3.7 analyses the versioning and history of changes in the process. This step is crucial in determining if this documentation includes details that could embody business strategies or product developments.

\noindent\textbf{PCRE Phase 4: Assess confidentiality of data elements (Stakeholders: Process Analyst)} Phase 4 scrutinizes the confidentiality of all process data elements ($\mapsto$ \textsl{Objective 3}). Question 4.1 starts with checking the atomicity of the data elements. This is important since atomic data elements, being the most granular level of data, often carry less risk of directly revealing sensitive information unless they are unique identifiers. Composite data elements, by contrast, encapsulate more detailed information due to their complex structure, posing a higher risk of revealing sensitive information. If data elements are composite, they are decomposed into sub-components until all of them are atomic. Decomposition is fundamental for assessing the confidentiality of the contained information. It also aids choosing appropriate anonymization techniques to minimize the loss of utility while maximizing privacy. 
For each of the atomic data elements, we pass through Questions 4.2 to 4.10. First, dependencies between data elements are determined (4.2), including process-based dependencies and functional dependencies. For determining data dependencies, a process model is required.
\begin{wrapfigure}{r}{0.43\textwidth}
  \begin{center}
    \includegraphics[width=0.42\textwidth]{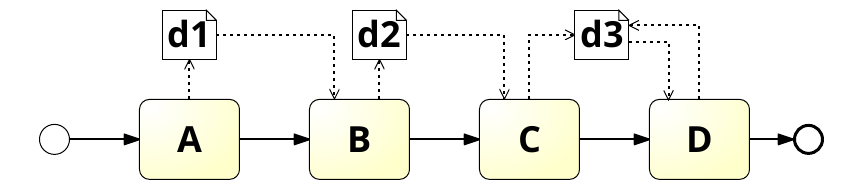}
  \end{center}
  \caption{Process With Data Dependencies}
   \label{fig:datadepex}
\end{wrapfigure}An example process-based data dependency is depicted in Fig. \ref{fig:datadepex} where data element d2 is written by task B after B reading d1. This analysis can also result in data dependency clusters where each of the contained data elements depend on each other, e.g., data elements d1, d2, d3 in Fig. \ref{fig:datadepex}. If one element within a cluster is considered sensitive or confidential, the other elements may also carry similar confidentiality concerns. Thus, data elements in the same dependency cluster should be anonymized to the same degree. Question 4.3 identifies the scope of the data element in terms of processes that access the data element, resulting in a list of process ids.
After identifying dependencies and scope, for each data element, its privacy and confidentiality requirements are assessed. Here, one process model is not sufficient, but several process models accessing a data element are to be analyzed ($\sim$ in Tab. \ref{tab:phase4}). Note that we assume that an atomic data element has either privacy or confidentiality requirements. Question 4.4 assigns a a risk of re-identifying an individual based on the the data element on a scale of 0--5 where 0 refers to `does not concern individuals', 1 to `no identification possible', 2 to `unspecific', 3 to `in combination with other elements', 4 to `strong delimitation possible', and 5 to `direct identification possible'. For these individual-related privacy concerns (rating $>1$), the required level of anonymization can be derived from the regulatory guidelines provided by the GDPR. A rating of 0 indicates that there are no privacy concerns, but there could be confidentiality concerns, resulting in assessment based on Questions 4.5 to 4.10 which balance the risk of publishing the data element (confidentiality loss) against its utility for the process analysis (utility loss). 

\begin{table}[htb!]
\begin{scriptsize}
 \centering\begin{tabular}{p{0.5cm}p{7.3cm}p{2.3cm}P{0.8cm}P{0.6cm}}
\textbf{ID} & \textbf{Question} & \textbf{Reply} & \textbf{Model} & \textbf{Log}\\\hline
4.1 & Is the data element atomic or composite? & atomic, composite & x & x \\
4.2 & On which other data elements does the data element depend? & list of data elements & x & \\
4.3 & Which (other) processes access the data element? & process ids & $\sim$ &  \\
4.4 & Can an individual be uniquely identified by the data element? & scale: 0--5  & x & x\\
4.5 & Does the data element contain information that should not be disclosed externally? & scale: 1--5 & x & x \\
4.6 & How likely are potential inferences? & scale: 1--5 & x & x\\
4.7 & How severe are the implications of the inference? & scale: 1--5 & x & x\\
4.8 & How critical is this data element for the functioning of the process? & scale: 1--5 & x & x\\
4.9 & What is the impact of the data element on decision-making in the process? & scale: 1--5 & x & $\sim$\\
4.10 & How frequently is the data element used in the process?  & scale: 1--5 & x & x\\\hline
\end{tabular}\caption{Checklist data elements for Phase 4}\label{tab:phase4}
\end{scriptsize}
\end{table}

Questions 4.5 to 4.7 assess the confidentiality risk when publishing the data element by estimating the severity and probability of an inference. The scale for a possible confidentiality risk (4.5) comprises 1: `none',
2: `unimportant sensitive information', 3: `sensitive information, but not critical', 4: `important sensitive information', and 5: `critical business secrets'. The scale for the likelihood of inference ranges from `very unlikely' (1), `unlikely' (2), `possible' (3), `likely' (4), to `very likely' (5). Finally, the severity of inference (4.7) is measured as 1: `none', 2: `minor negative', 3: `moderate negative', 4: considerable negative', and 5: `catastrophic'.  
Questions 4.8 to 4.10 evaluate the utility of each data element for the process and, consequently, its relevance for analysis. Question 4.8 estimates the criticality of the data element based a scale from 1--5, i.e., 1: `not critical', 2: `slightly critical', 3: `moderately critical', 4: `very critical', 5: `absolutely critical'. Question 4.9 rates the impact of a data element on decision making in the process based on 1: `no influence', 2: `minor influence', 3: `moderate influence', 4: `high influence', 5: `decisive'. Finally, Question 4.10 determines how frequent the data element occurs in the process, ranging from 1: `never', 2: `rarely', 3: `sometimes', 4: `often', to 5: `constantly'. 
The rationale behind Questions 4.8 to 4.10 is that an element that occurs in many conditions or is frequently retrieved, written, or modified is probably highly relevant to the overall process and thus implies a high utility loss upon anonymization. 

The provided ratings can be aggregated into different metrics, e.g., by averaging the ratings. Alternatively, the ratings can be weighted according to the trade-off between privacy and utility loss. Giving a higher weight to the questions centring around the risk imposed by publishing the data elements weighs the privacy aspects higher than the utility aspects, and vice versa. Finally, to account for the risk of a possible chain reaction of disclosures within a dependency cluster, the maximum rating of confidentiality among the data elements within the cluster can be taken.

It is advisable to give an example of a possible value  for each data element in order to ease the understanding of privacy-preserving requirements imposed by the data element for other stakeholders. Besides, the case may arrive that unused data elements exist within the process. These data elements can be suppressed entirely as they have no utility for analyzing the process and still contain confidential information. By the end of Phase 4, the organization obtains a list of all data elements present in the process, decomposed to an atomic level, and data dependency clusters which are mapped to a confidentiality rating.

\noindent\textbf{PCRE Phase 5: Consolidate results and prepare workshop (Artefact: guiding questions for consolidation and preparation)} Phase 5 aims at consolidating the results obtained in Phases 1--4 to prepare the workshop in Phase 6, and is carried out by the person responsible for executing PCRE. The consolidation can be supported by the guiding questions presented in Tab. \ref{tab:phase5}. These questions also aim at defining necessary anonymization actions to meet privacy and confidentiality requirements by, at the same time, minimizing the utility loss caused by them ($\mapsto$ \textsl{Objective 4}).

\begin{table}[htb!]
\begin{scriptsize}
 \centering\begin{tabular}{p{1cm}p{11cm}}
\textbf{ID} & \textbf{Question}\\\hline
5.1 & Which stakeholders should be included in the workshop? \\
5.2 & Which process elements have been classified as particularly confidential and why? In detail: processes/subprocesses (5.2.1), tasks (5.2.2), process parameters (5.3.4), data elements (5.2.4), endpoints (5.2.5), versioning/history of changes (5.2.6) \\
5.3 & Which privacy-preserving techniques, e.g., anonymization, are necessary to protect the process elements classified as confidential from disclosure? \\
5.4 & How can the necessary degree of privacy-preservation be achieved?\\
5.5 & Which KPIs or metrics can be calculated from the data elements and are classified as business secrets?\\
5.6 & What compromises must be made between the degree of privacy-preservation and the retention of data utility?\\\hline
\end{tabular}\caption{Template for guiding questions in Phase 5}\label{tab:phase5}
\end{scriptsize}
\end{table}
\vspace{-0.5cm}
\noindent\textbf{PCRE Phase 6: Conduct workshop with stakeholders (Stakeholders: Process Analyst, Management
Legal Compliance Team)} 
The workshop aims at obtaining approval for the minimal privacy-preserving actions such as anonymization, pseudonymization, or timestamp shifting while retaining the maximum possible utility. The workshop starts with revisiting the motivation behind data publication and the objectives of data analysis.
Subsequently, the results of Phase 3 on the process metadata elements and the data elements identified in Phase 4 are discussed. For each element, participants address the question: Given the specific element or class of dependent elements within the process log and the required level of privacy preservation, does its publication present any issues? In the course of this discussion, the approval of decision-makers within an organization, probably the management and legal compliance team, is obtained. Eventually, the predefined data analysis goals and assessment of the trade-off between privacy and utility are revisited.

\noindent\textbf{PCRE Phase 7: Consolidate workshop results } 
Phase 7 consolidates and documents the results of the workshop conducted in Phase 6 within an executive summary containing key decisions, goals, and objectives of data usage. To this end, the person responsible for conducting PCRE scrutinizes again all process elements contained in the process model and the process event logs w.r.t. the results obtained in Phases 1--6.

\section{Evaluation}
\label{sec:eval}

Development and evaluation of PCRE are conducted in an intertwined way, resembling an incremental development with  iteratively incorporating feedback:

\begin{itemize}
    
    \item \textbf{First round interviews:} Develop and refine the PCRE approach with altogether tree process analysts from two manufacturing companies and one process analyst from the medical domain.
    
    \item \textbf{Second round interviews:} We applied the PCRE for two scenarios in the same two manufacturing companies.
    
\end{itemize}

For the \textbf{first round interviews} we focused on process analysts that were available for the technical implementation and the continuous improvement of processes. During each phase, the process analysts reported experiences, advantages, and disadvantages. These insights were used to refine the PCRE approach. In order to complement the development of the PCRE method, we conducted another interview with a process analyst of a different manufacturing company on an ideation process and mix in experiences from discussions with a process analyst from the medical domain.

All first round interviews
were conducted online in an unstructured way between 10 Jan 2024 and 1 Feb 2024. 
One significant outcome was that the questions initially designed for Phase 3 did not indicate an assessment of each process metadata element. Consequently, the questions were enhanced to reflect the necessity to assess all process metadata, such as endpoints or task labels. 
Additionally, a more detailed description how to incorporate the results of Phase 3 into subsequent phases, especially Phase 6, was elaborated. 
During Phase 4, $21$ data elements were analysed, their atomicity documented, and their confidentiality assessed. However, exploring data dependencies proved to be cumbersome due to the complex interconnections among some elements. First, the functional dependencies are deeply intertwined and complex to document. Their multiplicity presents significant challenges in assessment. Secondly, there are additional considerations when evaluating confidentiality: certain data elements only reveal confidential information when analysed in combination with other data elements. Considering, for example, timestamps and product IDs individually does not comprise confidential information. However, when combined, they could facilitate the derivation of sensitive information. This observation is reinforced by discussions with a process analyst from the medical domain for combinations of timestamps and patient IDs. Another insight was to document example values and additional notes when evaluating the confidentiality of the data elements to enhance understandability. 
Overall, meaningful insights could be gathered in Phase 4, facilitating the preparation of the upcoming workshop. An additional benefit of PCRE was that redundant and deprecated data elements were identified by documenting all process elements. Removing these data elements enhances process efficiency, documentation quality, and overall understandability of the process.

\noindent\textbf{Second round interviews:} We repeated all interviews for two concrete (different from first round interviews) processes/scenarios with the updated PCRE Phases and questions, between 4 and 8 months later. For the first manufacturing company we relied on the same process analysts, while for the second manufacturing company we utilized a second available process analyst. The preparations, interviews, and consolidation (Phase 1 to 5) could be done in about 5 hours for each company.

Then, for both companies workshops were conducted (Phase 6). The participants for the first company included two process analysts, a division manager filling the management role, a team leader of the manufacturing department, bridging the gap between employees and management, and a deputy team lead involved in the technical implementation of the process. In the following, we summarize the main results of the workshop for company a. First, different stakeholders have different interests and perspectives, fundamentally influencing their focus and concerns regarding data analysis objectives. The upper management primarily focuses on long-term effects and strategic planning aspects such as monthly evaluations and growth rates. In contrast, the interests of persons directly responsible for the day-to-day operations require short-term information such as current capacity utilization and real-time problem reporting. 
Second, the stakeholders agree on the value that could potentially be generated by exploring process data. An important question is how the manufacturing company can access the results of the process data analysis. All process endpoints are identified as confidential since they expose the internal network topology. 
Although only selected tasks and task labels are determined as confidential, all tasks and task labels throughout the event log need to be anonymized as they are considered too ambiguous, hence leaving a broad scope for interpretation, possibly leading to false conclusions and thus reducing their utility for data analysis. Therefore, both, tasks and task labels will be generalized. 
Third, the publication of data elements is considered difficult, as the perceived risks often outweigh the benefits. Moreover, the data elements considered publishable typically contribute minimally to process insights, offering reduced utility for process analysis. A single trace within the manufacturing process relates to the manufacturing of one product instance. This poses an important differentiation: Confidentiality concerns increase if several traces are published, because additional metrics, such as production volumes or process changes over time, will be recognizable. For example, compliance with legal requirements and adaptability to legal changes are indicated, and periods with low manufacturing performance or quality problems can be identified. Furthermore, a data element relevance can also span multiple instances, traces, or processes. Especially when process data of various (sub) processes is published, a data element of one process could be used to disclose information from another process. 

For the second company, the 3 participant included a team lead responsible for KPI development, and 2 process analyst, one responsible for technical implementation, and one responsible for overseeing the evolution of the process. This workshop focused on how to created different data packages for internal use, with different analysis goals. It was agreed, that because of the nature of the process, for making the data publicly available, most data had to removed, only leaving the duration of process instances and frequency and occurrence of certain repeated tasks as valid analysis tasks.

As conclusion, conducting PCRE enabled the systematic discussion of all process-related information with different stakeholders, weighing the benefits of process analysis results with privacy and confidentiality threats. Moreover, PCRE resulted in cleaning process data, i.e., deprecated data element, resulting in a more streamlined and efficient process. In this case, the executable process model was used as basis for PCRE assessment, resulting in an assessment of the generated event logs. In the discussion with process analysts from the medical domain, the basis are process event logs. Here, the assessment helped to identify limited process analysis results caused by too strict privacy-preserving measures w.r.t. timestamps in the log, i.e.,  providing time distances between events instead of timestamps resulting in the non-applicability of process discovery. Slightly adapting this privacy-preserving measure to shifting timestamps consistently for a selection of process instances enables richer process analysis results without infringing privacy and confidentiality.   

\noindent\textbf{Validation interview on ideation process:} 
The controlled experiment is conducted based on an ideation process, reflecting a data-driven approach to document and evaluate new business ideas, projects, or products within an organization. The process model is inspired by a real-world project and implemented using the Cloud Process Execution Engine CPEE\footnote{\url{https://cpee.org}, last access: 2025-01-25}. This facilitates the definition of data elements which can be examined before the process comes into production (process testing). The process consists of $12$ activities and $18$ data elements. Moreover, three endpoints are defined, and $12$ process attributes (parameters) are set. Most data elements are links to files, e.g., large JSON files, and hence not atomic, making the assessment more complex. The endpoints are predefined and lead to HTML forms usable for data input or PHP scripts, e.g., for sending an email. The process parameters comprise the unique identifier of the process model and personal identifiers such as names of the creator and author.

The process model itself is rather generic and, therefore, not confidential. In contrast, data produced throughout the process is highly sensitive, since new products or business ideas are of the utmost value for a company. 

For the controlled experiment, a research assistant took the position of the process analyst, given that she was responsible for the implementation of the process. The role of a representative of the management and the role of a member of the legal compliance team were taken by two other research assistants. An author of this work acted as external consultant responsible for executing PCRE and the preparation of each phase. The Excel file used to conduct the PCRE for the ideation process can be found on here\footnote{\url{https://github.com/fahaerte/privacy_confidentiality_in_process_mining/tree/main/pcre_approach}, last access: 2025-01-25}. In the following the results of each phase are described in detail.

\noindent\textsl{Phase 1:} 
This phase was conducted as an interview with the process analyst. The results (cf. Tab. \ref{tab:phase1}) are as follows: (1.1) a process model is available and implemented in CPEE and an event log is available; (1.2) the process has not yet been applied in practice; therefore, data has not been used for analysis; (1.3)
management and the R\&D department are interested; (1.4) management, legal compliance team, process analyst.

\noindent\textsl{Phase 2:}
The second phase was conducted as an interview with the process analyst. The results (cf. Tab. \ref{tab:phase2}) are as follows: (2.1) the process has not yet been applied in practice; (2.2) performance analysis, process discovery, conformance checking; detecting evolution and deviations of the process; (2.3) duration of single process steps, the conversion rate of developed ideas, an inspection of the granularity of the process steps, identification of bottlenecks, the effect of the involvement of certain persons throughout the ideation process, the effect of the involvement of a certain amount of people throughout the process, the necessity of single process steps; (2.4) metrics are currently not calculated (2.5) metrics for 2.3; (2.6) no insights can be identified.

\noindent\textsl{Phase 3} was conducted in an interview format with the process analyst. The results (cf. Tab. \ref{tab:phase3}) are as follows: (3.1) no subprocesses; (3.2) 
process name is not considered confidential; (3.3) Currently tasks do not contain confidential information, and the code in the tasks only saves results in the data elements. In the future, the code may also contain business secrets and a re-assessment is necessary upon modification of the process; (3.4)
task labels are not confidential as the process is generic; (3.5) process parameters ``creator'' and ``author'' contain personal data and must be suppressed; ``design\_dir''  is considered sensitive and must be suppressed. The remaining parameters are not considered sensitive; (3.6) all endpoints are confidential and must be suppressed; (3.7) The change history is possibly contained in the event log and is currently not considered sensitive, but this could change when the process is modified.

\begin{figure}[htb!]
    \centering
    \includegraphics[width=0.8\textwidth]{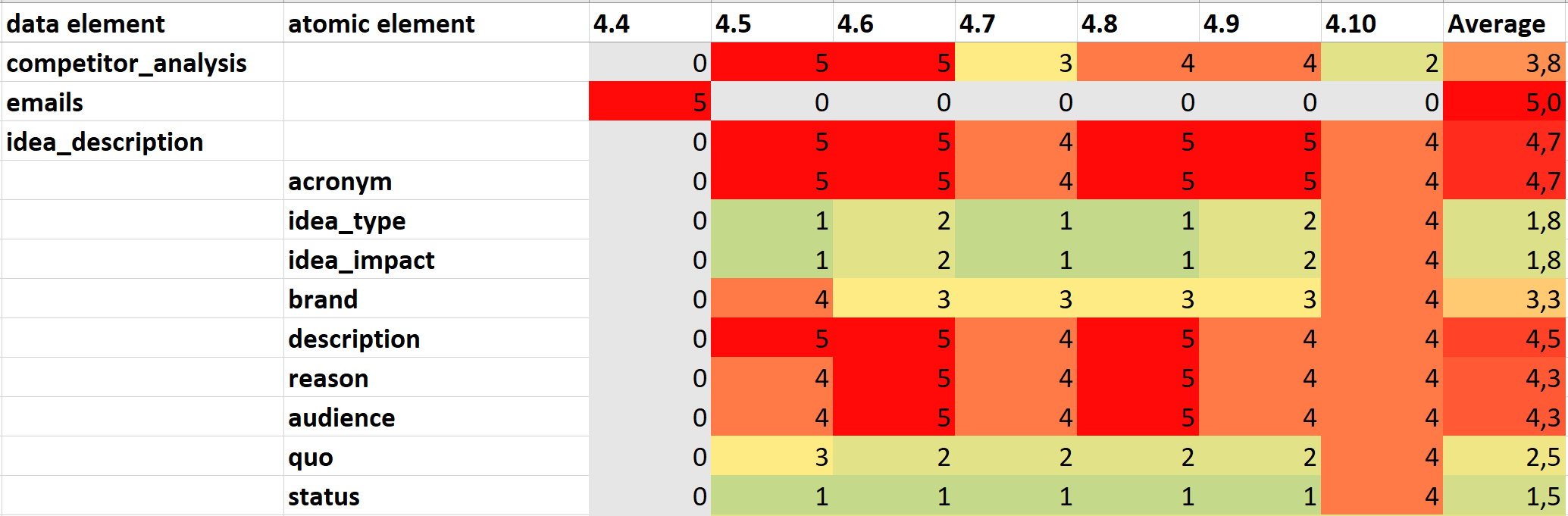}
     \caption{Excerpt of Results of Phase 4 in Controlled Experiment}
    \label{fig:experimentphase4}
\end{figure}

\noindent\textsl{Phase 4} was conducted together with the process analyst by filling the checklist for assessing the confidentiality of each data element  (cf. Tab. \ref{tab:phase4}). We constructed the ``Smart Lighting Tool'' product idea, which turns out useful to have meaningful example values for each data element. A selection of results are presented in the following (see Fig. \ref{fig:experimentphase4}). The full results can be found in the github repo. 

(4.1) Some data elements are atomic, e.g., `competitor\_analysis' with an example value of \texttt{https://cpee.org/modified/test/link}. Other data elements are structured as JSON elements and were decomposed into their atomic parts. One example is data element `idea\_description' which was decomposed into its atomic parts, e.g., `acronym', `idea\_type'. (4.2) For analysing the dependencies, it is essential to consider the granularity and interrelations of data elements, especially when encapsulated within complex structures such as JSON objects. Within the JSON object `idea\_description', for example, atomic components such as `acronym' and `description' are intrinsically linked since the acronym is further detailed in the description (functional dependency). Data elements `mission\_one\_pager' and `problem\_canvas' interlink concepts such as `mission', `goals', and the problem statement contained in `problem' and collectively create a narrative about the product purpose and the market needs. Finally, the approval status of the `gate\_decision' element depends on assessing all previous elements. The justifications provided within `gate\_decision' likely reflect the entire process data. (4.3) There are not other processes. 

(4.4) Data element `idea\_description', for example, does not deal with individual data (rating 0). (4.5) It contains critical business secrets (rating 5). (4.6) A potential inference is very likely (rating 5), (4.7) having considerable negative implications (rating 4). (4.8) Since the whole process is centered around evaluating the new idea, the data element is absolutely critical for the entire process (rating 5). (4.9) It has decisive impact on the decision-making (rating 5). (4.10) Lastly, the data element is used often throughout the process (rating 4). This results in an overall confidentiality rating of 4.7 and, therefore, makes suppression of the data element inevitable since a modification will highly reduce the data utility, making further reasoning redundant. 
It becomes sufficient to anonymize some parts when decomposing `idea\_description' into atomic parts. For example, atomic parts acronym' and `description' require thoughtful anonymization. Parts `idea\_type' and `status' are rated as less confidential. These elements may remain unmodified in the process data, preserving overall data utility.

\noindent\textsl{Phase 5:} The answers collected in Phase 1--4 were reformulated, structured, and detailed in this phase to ensure comprehensiveness and understandability.

\noindent\textsl{Phase 6:}
The workshop simulation was carried out collaboratively with the process analyst, management, and legal compliance team. The KPI on the effect of the involvement of certain persons in the ideation process was changed to the analysis of the effects of the involvement of certain roles due to concerns of raised by the legal compliance team. The other results collected in Phase 2 were confirmed. The proposed anonymization methods for tasks, task labels, endpoints, and process parameters, were reviewed and accepted without additional modifications (Phase 3). The discourse on Phase 4 results revealed significant concerns primarily from management regarding the potential exposure of sensitive business strategies through the publication of internal documents in, e.g., `competitor\_analysis', necessitating suppression as initially proposed. Similarly, the legal compliance team advocated suppressing the email addresses in the data element `emails' to protect individuals. Particularly challenging was the discussion around the atomic components of `idea\_description'. The management argued for suppression of the atomic components `acronym', `description', and `reason' due to the risk of unveiling innovative product ideas and strategic objectives. Moreover, the `idea\_type' and `idea\_impact', despite containing generic values, were viewed as potential leaks of strategic objectives. Thus, both elements were designated for generalization. Elements `audience' and `brand' were addressed by replacing specific identifiers with generic pseudonyms, which all stakeholders agreed was sufficient by contrast to the initially proposed anonymization technique of suppression. The resulting anonymization techniques are:
\begin{itemize}
    \item \textsl{No anonymization:} tasks and task labels
    \item \textsl{Suppression:} all endpoints; process parameters creator, author, design\_dir; data elements competitor\_analysis, portfolio\_analysis, emails, \\idea\_description, idea\_description[acronym], idea\_description[description],  idea\_description[reason]
    \item \textsl{Generalization:} idea\_description[idea\_type], idea\_description[idea\_impact], idea\_description[brand],  idea\_description[audience]
\end{itemize}

\noindent\textsl{Phase 7} yielded the following executive summary: 

\noindent$\bullet$ Data related to individuals in the company such as names are identifiable in the event log and hence will be suppressed to prevent any possibility of linking names to roles or other personal identifiers, thereby complying with the GDPR. \\
\noindent$\bullet$ Three process parameters and all endpoints contain confidential or private information and will be suppressed upon publishing.\\
\noindent$\bullet$ The event log contains proprietary information, including new product ideas and internal documents intended for confidential internal use. Additionally, insights into the company’s strategic objectives, such as the product mix or market focus, are classified as business secrets, needing protection from disclosure. \\
\noindent$\bullet$ The process data yields metrics and KPIs mainly from text-based evaluations and time-related analysis, such as the duration of process steps and the identification of bottlenecks. These KPIs are not confidential and remain computable despite anonymization, preserving their utility for performance analysis.\\
\noindent$\bullet$ The defined analytical objectives include examining the duration of a single process steps, the conversion rate of developed ideas, the sufficiency of the granularity of process steps, identification of bottlenecks, the effect of the involvement of certain roles, and the effect of the involvement of a certain amount of people throughout the Ideation Process.\\
\noindent$\bullet$ All data analysis goals can be met despite the anonymization measures.\\
\noindent$\bullet$ With the anonymization steps in place, privacy and confidentiality concerns are sufficiently addressed. The remaining risk of the potential deduction of business strategies or the unveiling of new product ideas has been evaluated and tackled with the outlined anonymization steps. The trade-off between privacy and utility has been examined, ensuring that the value of the data for analysis justifies any minimal loss in utility due to anonymization.

\section{Related Work}
\label{sec:relwor}

Existing literature provides methods and techniques for the elicitation, modeling, and verification of security and privacy requirements in the context of business processes, e.g., \cite{DBLP:journals/bise/MatuleviciusNS18,DBLP:journals/infsof/LeitnerR14} as well as for augmenting process models with privacy-enhancing technologies \cite{DBLP:journals/istr/BakhtinaMS23,DBLP:journals/sttt/DumasGJLMPPPTTY22}. 

Another research stream refers to privacy-preserving methods in process mining.
Existing approaches in this stream can be distinguished into grouped-based privacy models, indistinguishability-based models, and confidentiality frameworks \cite{DBLP:journals/tmis/ElkoumyFSKMVRW22}. Grouped-based privacy models rely on the concept of k-anonymity \cite{DBLP:journals/ijufks/Sweene02}, extended by the notions of l-diversity \cite{machanavajjhala2007diversity}, and t-closeness \cite{li2006t}. They comprise the TLKC model \cite{DBLP:conf/rcis/RafieiWA20}, PRETSA \cite{DBLP:conf/icpm/Fahrenkrog-Petersen19}, and \cite{batista2022privacy}. Indistinguishability-based models utilize differential privacy to ensure that two versions of a data set cannot be differentiated up to a specified level. Approaches in these group are  \cite{DBLP:journals/bise/MannhardtKBWM19,DBLP:conf/bpm/Fahrenkrog-Petersen20,DBLP:conf/bpm/RafieiA19,rafiei2022travas,kabierski2023hiding,elkoumy2022libra}. Finally, confidentiality frameworks utilize encryption methods and access control regulations to restrict access to confidential data or results from data analysis to authorized parties  \cite{DBLP:books/sp/22/Mannhardt22,DBLP:journals/tmis/ElkoumyFSKMVRW22,DBLP:conf/simpda/RafieiWA19}. 
Moreover, several tools have been demonstrated for privacy-preserving process mining. Examples include ELPaaS \cite{DBLP:conf/bpm/0006FKMAW19} that supports the generation of an anonymized event log satisfying privacy metrics, and mining techniques resulting in desired privacy guarantees as well as  ShareProm \cite{DBLP:conf/bpm/ElkoumyFDLPW20} enabling a secure multi-party computation framework. 

\noindent\textsl{Research gap:} Process mining related techniques do not consider confidentiality beyond information of individuals and have not yet balanced risk  and utility \cite{DBLP:journals/tmis/ElkoumyFSKMVRW22}. These gaps can be addressed by PCRE. Nonetheless, existing approaches can be applied in a complementary way by dealing with identified privacy concerns. This also holds true for approaches for eliciting security requirements and equipping processes with privacy-enhancing technologies

\section{Conclusion and Outlook}
\label{sec:conclusion}
In summary, the PCRE method features seven phases with questionnaires, checklists, ratings, and guidelines, to systematically assess privacy and
confidentiality requirements of process data within an organization. It is generic and broadly applicable across various domains, incorporating quantitative and qualitative evaluations of the confidentiality of process
elements. Weights can be defined to adjust PCRE to the domain and specific needs of the organization. Moreover, PCRE facilitates the comprehensive documentation of processes and  encourages a multidisciplinary dialogue among diverse organizational stakeholders. Through
such a collaboration, stakeholders critically analyse event logs for process analysis and determine
relevant metrics or KPIs. Therefore, aligning privacy-preserving objectives with Process Mining
and data analysis capabilities is ensured by design.


\noindent\textbf{Limitations and future work:} First, in order to collect diverse interests and perspectives on data analysis objectives, the PCRE methods requires multiple stakeholders to assess the current state of data analysis within the organization. This might become difficult, especially for external consultants, and result in resource and communication overhead. Second, the availability of the process data constitutes a potential challenge. In this work, the processes are implemented in a process engine, resulting in well-specified process parameters, data element, and endpoints. In practice, process data might be available at different levels of quality, directly influencing the complexity of conducting PCRE. Third, the number and complexity of data dependencies might also increase the complexity of the analysis, especially at the presence of multiple sub-processes. Additionally, detailed guidelines for mapping
confidentiality assessments to specific anonymization techniques could streamline the anonymization
process. In future work, we will apply PCRE in other application domains, e.g., health care, in order to evolve the method into a robust framework.

\end{document}